\documentclass[a4paper,12pt]{article}
\title{MSM_pseudo_CS}
\usepackage{amssymb, amsmath, amsthm, mathtools, color}
\usepackage{booktabs}       % professional-quality tables
\usepackage{multirow}
\usepackage{multicol}        % used for the two-column index
\usepackage{longtable}
\usepackage{array}
\usepackage{float}
\usepackage{subfigure}
\usepackage{subfig}
\usepackage{epsfig}
\usepackage{url}
\usepackage{graphicx}
\usepackage{csquotes}
\usepackage{enumerate}
\usepackage{enumitem}
\usepackage{amsmath}
\usepackage{listings}
\usepackage{color}
\usepackage{physics}
\usepackage{listings}
\usepackage{comment}
\usepackage[title]{appendix}
\usepackage[numbers]{natbib}
\usepackage{comment}
\usepackage[figuresright]{rotating}
\usepackage{tikz}
\usetikzlibrary{arrows.meta}
\tikzset{>={Latex[width=3mm,length=3mm]}}
\usetikzlibrary{positioning}
\usepackage{lineno}

\title{Nonparametric estimation of a state entry time distribution conditional on a ``past'' state occupation in a progressive multistate model with current status data}
\author{Samuel Anyaso-Samuel${^1}$ and Somnath Datta${^2}$ \\ \small ${^1}$Division of Cancer Epidemiology and Genetics, National Cancer Institute, Rockville, MD, USA\\ \small ${^2}$Department of Biostatistics, University of Florida, Gainesville, FL, USA}
\date{}

\begin{document}
\maketitle

\begin{abstract}
Case-I interval-censored (current status) data from multistate systems are often encountered in biomedical and epidemiological studies. In this article, we focus on the problem of estimating state entry distribution and occupation probabilities, contingent on a preceding state occupation. This endeavor is particularly complex owing to the inherent challenge of the unavailability of directly observed counts of individuals at risk of transitioning from a state, due to severe interval censoring. We propose two nonparametric approaches, one using the fractional at-risk set approach recently adopted in the right-censoring framework and the other a new estimator based on the ratio of marginal state occupation probabilities. Both estimation approaches utilize innovative applications of concepts from the competing risks paradigm. The finite-sample behavior of the proposed estimators is studied via extensive simulation studies where we show that the estimators based on severely censored current status data have good performance when compared with those based on complete data. We demonstrate the application of the two methods to analyze data from patients diagnosed with breast cancer.
\end{abstract}

\clearpage
\section{Introduction}\label{sec:intro}
Multistate models are a foundational framework for analyzing disease progression with multivariate time-to-event data in biomedical research. Individuals typically start in an initial state and move through intermediate disease states before possibly entering an absorbing state. In many settings, however, continuous follow-up is infeasible and the process is observed only at a single random inspection time per individual (``current status'' or case-I interval censoring), so that for each individual one records only the state occupied at that inspection \citep{sun1993,jewell2003,sun2006}. 

In progressive systems with a directed tree structure, a clinically pertinent question is: \emph{what is the probability of ever occupying a given state $k$, conditional on a prior visit to state $j$}? We denote this marginal (population-level) probability by $\Psi_{k\mid j}$, and its time-specific analogue by $\Psi_{k\mid j}(t)$. These quantities summarize the proportion of individuals who, among those who ever visit $j$, ultimately (or by time $t$) occupy $k$ on the unique path from the root to $k$. Estimation of such marginal-conditional probabilities can inform prognosis and planning after intermediate milestones. For example, risk of distant metastasis after loco-regional recurrence in oncology or risk of a second remission after a first remission in hematologic malignancies \citep{putter2006,klein2002,vicini2003,murata2023}. 

For right-censored event-history data from progressive systems, estimators of state-occupation probabilities conditional on prior state occupation have recently appeared \citep{yang2023}. Under current-status observation, the problem is tougher: one never observes the future path after the single inspection, and even transitions out of nonabsorbing states are only partially observed. This paper develops two nonparametric estimators of $\Psi_{k\mid j}$ and $\Psi_{k\mid j}(t)$ for progressive multistate models observed under current status, without imposing the Markov property. The first estimator builds fractional at-risk sets by weighting each individual’s contribution to the at-risk set of transitioning out of state $j$ by the probability that the individual would ever reach $j$, adapting ideas of fractional weighting from Datta and Satten\cite{datta2000} and connections to Yang et al. \cite{yang2023} The second estimator exploits the identity of $\Psi_{k\mid j}$ as a ratio of marginal occupation probabilities in tree-structured systems, leading to a product-limit–type construction based on multistate state-occupation estimators \citep{datta2006}. We also derive estimators of the distribution of the entry time into $k$ given prior occupation of $j$.

The rest of the paper is organized as follows. Section~\ref{sec:design} formalizes the single-inspection multistate design and gives biomedical examples where such data arise. Sections~\ref{sec: FRE} and~\ref{sec: PLE} develop the fractional-risk and product-limit estimators, respectively. In Section \ref{sec: confint}, we describe the estimation of pointwise confidence intervals for the estimators based on both approaches. Section \ref{sec: pseudo} discusses testing of covariate effects via pseudo-value regression. Detailed simulation studies to evaluate the performance of the proposed estimators are presented in Section \ref{sec: sim}. In Section \ref{sec: eortc}, we illustrate the application of both estimation methods in analyzing data from a breast cancer study. The paper concludes with a discussion in Section \ref{sec: discussion}.

\section{Study Design: Single-Inspection Multistate Sampling}\label{sec:design}

\subsection{What is observed under multistate current status?}
Consider a progressive multistate process $\{S_i(t): t\ge 0\}$ on a directed tree $\mathcal{S}=\{0,1,\ldots,Q\}$ with unique root-to-node paths. Under \emph{current-status} observation, subject $i$ is inspected once at a random time $C_i$ and only the pair $\{C_i, S_i(C_i)\}$ is recorded. The inspection times $\{C_i\}$ are independent of the underlying multistate process, based on the random censoring assumption. No transition times or future trajectories after $C_i$ are observed. Thus, the data consist of a single cross-sectional snapshot of the state occupancy across heterogeneous times since a common origin (e.g., time since diagnosis, surgery, transplantation, or age).

This observation scheme is stricter than right-censoring but is practically motivated whenever repeated follow-up is infeasible, expensive, or ethically constrained (e.g., one-time biospecimen collection). 

\subsection{How common is the design in practice?}
We acknowledge that multistate current-status data are not as ubiquitous as right-censored multistate cohorts in public repositories. Nevertheless, single-inspection designs arise in several biomedical domains, and importantly, they produce multistate current-status information:

\begin{enumerate}[label=(\alph*)]
\item \textbf{Two-state (classical) current status in screening and serosurveys.} Cross-sectional surveys that ascertain whether an event has occurred by age or time-since-exposure are textbook examples of current-status designs in epidemiology and screening \citep{sun2006}. 

\item \textbf{Progressive infection staging from a single biospecimen.} 
\emph{HIV recency assays} classify cross-sectional specimens into ``recent'' versus ``long-standing'' infection to support incidence surveillance; individuals can be viewed as moving from susceptible $\to$ recent infection $\to$ established infection along a progressive path observed at one time point \citep{kassanjee2017cross,facente2022use}. 

\item \textbf{Serology-defined multistate profiles with branching.} 
\emph{Hepatitis B} screening panels (HBsAg, anti-HBc, anti-HBs) classify single-visit laboratory results into mutually exclusive clinical states: susceptible, immune due to vaccination, immune from past infection, chronic infection; forming a progressive tree with extended branches observable at one inspection (e.g., by age) \citep{schillie2018prevention}. 

\item \textbf{Staged chronic disease from a single clinical exam.} 
\emph{Periodontitis} is routinely staged in cross-sectional surveillance, producing ordinal disease states (Stage I–IV) on a progressive path observed at one dental exam; recent methodological work is explicitly motivated by cross-sectional multistate current-status periodontal data \citep{anyaso2023}. 
\emph{Diabetic retinopathy} screening programs similarly classify a single fundus image into graded states used for referral and surveillance, again yielding a progressive multistate snapshot \citep{nhs2012gradingcriteria}. 
\end{enumerate}

Although multistate current-status datasets with extended branching are less common in public repositories than right-censored cohorts, the design often arises in practice as indicated above. Because publicly available examples with rich branching are scarce, for illustration, researchers often emulate a single-inspection design from an existing right-censored dataset by masking follow-up to one random inspection per subject, consistent with the random inspection-time assumption. 

\subsection{Notation and Convention}\label{sec: notation}
For each state $k \in \mathcal{S}$, let $\mathcal{P}(k) = \{0 = s_1 \rightarrow s_2 \rightarrow \cdots \rightarrow s_{m} = k\}$ denote the unique directed path from the root node $0$ to state $k$, where $m$ is the number of steps required to reach $k$. A state $j$ is said to lie on the path to $k$ if $j \in \mathcal{P}(k)$ and $j \neq k$. 

{\color{black}Let $U_{i,jj'}$ denote the (unobserved) transition time at which individual $i$ transitions from $j$ to $j'$.} Also, let $X_{ij}$ be the (unobserved) indicator that individual $i$ would eventually visit state $j$. The main estimands of interest, $\Psi_{k \mid j}$ and its time-specific version $\Psi_{k \mid j}(t)$, are defined for states $j \in \mathcal{P}(k)$ and represent, respectively, the marginal probability of ever occupying state $k$ given a prior visit to $j$, and the probability of occupying state $k$ by time $t$ given prior occupation of $j$. These quantities are defined at the population level and are identifiable from single-inspection data under the assumptions described.

\section{Estimators based on fractional at-risk sets}\label{sec: FRE}
\subsection{An illustration using a five-state model}\label{sec: 5state}
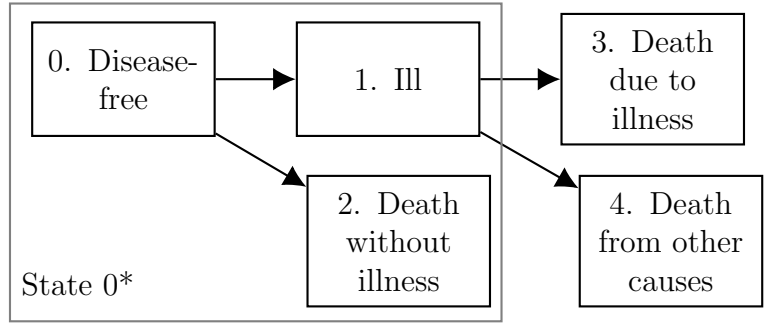
\begin{figure}
\centering
\begin{tikzpicture}[node distance={3.5cm}, thick, state/.style = {draw, rectangle, align=center,inner sep=0.5em, text width=2cm,minimum height=1.5cm}] 
\node[state] (0) {0. Disease-free}; 
\node[state] (1) [right of=0] {1. Ill}; 
\node[state] (2) [below right=0.5cm and 1.2cm of 0] {2. Death without illness}; 
\node[state] (3) [right of=1] {3. Death due to illness}; 
\node[state] (4) [below right=0.5cm and 1.3cm of 1] {4. Death from other causes}; 
\draw[->] (0) -- (1); 
\draw[->] (0) -- (2); 
\draw[->] (1) -- (3); 
\draw[->] (1) -- (4); 
% Boxing and labelling noise shapers
\draw [color=gray,thick](-1.5,-3.2) rectangle (5,1);
\node at (-1.5,-2.7) [above=5mm, right=0mm] {State 0*};
\end{tikzpicture} 
\caption{A five-state illness-death model.} \label{fig: illness-death}
\end{figure}

To provide context for the aim of this study, let's examine the progressive five-state illness-death model depicted in Figure \ref{fig: illness-death}. Here, state occupation would be determined through medical examinations. Our focus centers on estimating the probability of occupying state 3, given a prior visit to state 1 over an individual's lifespan. We denote this conditional occupation probability as $\Psi_{3|1}$. Estimating $\Psi_{3|1}$ poses a challenge due to incomplete observation of transition times. With current status data, we only have access to information about the state a patient occupies at a random inspection time.

In the typical current status framework, $X_{ij}$ will likely be unknown, making inference on a later state difficult. For instance, if the $i$th individual was observed to have $S_i(C_i) = 0$, then $X_{i1}$ will be unobserved. Suppose the $i$th individual transitioned to state 1, they will be eligible for the competing risks (state 3 and state 4), but we would not know this because $X_{i1}$ is not observed. To overcome the difficulty in estimating $\Psi_{3|1}$, first, we define a new state  $0^*$ which is constructed by combining states 0, 1, and 2 (see Figure \ref{fig: illness-death}). Note that not everyone (i.e., those who transitioned to state 2) in state  $0^*$ will be subject to the competing risks. 

Next, we compute the fractional observation, denoted as $\phi_{i1}(C_i, S_i(C_i))$, representing the contribution of the $i$th individual to the ``at-risk'' set for transitioning out of state 1. This fractional observation estimates the probability of the $i$th individual eventually reaching state 1 from state 0. Computing $\phi_{i1}(C_i, S_i(C_i))$ is straightforward when $S_i(C_i) \neq 0$. For $S_i(C_i) = 0$, we compute $\phi_{i1}(C_i, S_i(C_i))$ as the transition probability of moving from state 0 to state 1, that is, $\phi_{i1}(C_i, S_i(C_i)) = {P}_{01}(C_i, \infty)$. Using the Aalen-Johansen formula for transition probability, we estimate $\phi_{i1}(C_i, S_i(C_i))$ by the following expression
\begin{align}
    \widehat{\phi}_{i1}(C_i, S_i(C_i)) = \int_{C_i}^{\infty} \left\{ \prod_{(C_i,u)} \left[1 - \frac{d\widehat{N}_{01}(v) + d\widehat{N}_{02}(v)}{\widehat{Y}_{0}(v)}  \right] \right\} \frac{d\widehat{N}_{01}(u)}{\widehat{Y}_{0}(u)}.
    \label{eqn: phi1}
\end{align}
Here, $\widehat{N}_{\ell \ell'}(t)$ is the marginal estimator of the counting process $N_{\ell \ell'}(t)$, which counts the transitions from state $\ell$ to state $\ell'$ in the time interval $[0,t)$, and $\widehat{Y}_{\ell}(t)$ is the estimator of the at-risk process $Y_{\ell}(t)$, representing the number of individuals at risk of transitioning out of state $\ell$ at time $t$. The details of the construction of these estimators via nonparametric regression are provided in Appendix \ref{sec: datta2006} (\textit{ref:} (\ref{eqn: CPest})-(\ref{eqn: ARest})).

Importantly, the calculation of (\ref{eqn: phi1}) does not rely on the Markov assumption, as it only concerns the transition from the root node (state 0). If $S_i(C_i) = 2$, then $\widehat{\phi}_{i1}(C_i, S_i(C_i)) = 0$. For $S_i(C_i) \in \{1,3,4\}$, indicating that the individual followed the $0 \rightarrow 1$ path, $\widehat{\phi}_{i1}(C_i, S_i(C_i)) = 1$.

To obtain the estimator of $\Psi_{3|1}$, we introduce a modified multistate system that includes state 0* as the initial stage, leading to either state 3 or state 4 (see Figure \ref{fig: illness-death}). Applying the Aalen-Johansen formula, commonly used in competing risks calculations, to this modified multistate system yields an expression that serves as the estimator for eventually reaching state 3:
\begin{align}
    \widehat{\Psi}^{[1]}_{3|1} =\int^\infty_0 \left\{ \prod_{(0,u)} \left[1 - \frac{d\widehat{N}_{3|1}(v) + d\widehat{N}_{4|1}(v)}{\widehat{Y}_{0^*|1}(v)}  \right] \right\} \frac{d\widehat{N}_{3|1}(u)}{\widehat{Y}_{0^*|1}(u)}.
    \label{eqn: cond}
\end{align}
Here the superscript $[1]$ is used to distinguish this estimator from the alternate estimator discussed in Section \ref{sec: PLE}. To conduct the nonparametric regression, let $K>0$ represent a log-concave (differentiable) density, and let $0 < h = h(n) \downarrow 0$ be a bandwidth sequence. The estimators of the counting processes and the at-risk set in (\ref{eqn: cond}) are defined by
\begin{align*}
    \widehat{N}_{3|1}(t) &=  \sum^{n}_{i=1} \left\{\frac{\widehat{N}^P_{13}(t) K_h(C_{i} - t)}{\sum^{n}_{i=1} K_h(C_{i} - t)} \right\}, \\
    \widehat{N}_{4|1}(t) &=  \sum^{n}_{i=1} \left\{\frac{\widehat{N}^P_{14}(t) K_h(C_{i} - t)}{\sum^{n}_{i=1} K_h(C_{i} - t)} \right\}, \\
     \widehat{Y}_{0^*|1}(t) &= \sum^{n}_{i=1} \left\{\widehat{\phi}_{i1}(C_i, S_i(C_i)) \times \frac{\mathcal{I}(S_{i}(C_{i}) \in \{0,1,2\}) \times K_h(C_{i} - t)}{n^{-1}\sum^{n}_{i=1} K_h(C_{i} - t)} \right\}, 
\end{align*}
where $n^{-1}\widehat{N}^P_{jj'}(\cdot)$ is a step function for taking values $n^{-1}\widehat{N}^P_{jj'}(C_{(i)}) = R_{i}$, that minimizes the sum of squares $\sum_{i=1}^{n} \Big\{R_{i} - \mathcal{I}(U_{[i],jj'} \leq C_{(i)})\Big\}^{2}$ subject to $R_{1} \leq ... \leq R_{n}$, and $[i]$ corresponds to the index such that $C_{[i]}$ equals the $i$th-order statistic $C_{(i)}$. Subsequently, to obtain, $\widehat{N}_{j'|j}(\cdot)$, similar to the approach used in constructing (\ref{eqn: CPest}), we use kernel smoothing to remove the long flat part from $\widehat{N}^P_{jj'}(\cdot)$. Note, the estimator, $\widehat{Y}_{0^*|1}(t)$ is formulated by the estimate of the fractional weight, $\widehat{\phi}_{i1}(C_i, S_i(C_i))$. Here, $\widehat{Y}_{0^*|1}(t)$ estimates the number of persons at risk of transitioning out of the artificial state $0^*$ at time $t$.

Additionally, the estimation of the entry time distribution into state 3, given a prior visit to state 1, is expressed as follows:
\begin{align*}
    \widehat{F}_{3|1}(t) = \frac{\widehat{\Psi}^{[1]}_{3|1}(t)}{\widehat{\Psi}^{[1]}_{3|1}}.
\end{align*}
Here, the estimation of the subdistribution function ${\Psi}^{[1]}_{3|1}(t)$ is achieved by constraining the integral's bounds as defined in the expression provided in (\ref{eqn: cond}). Therefore,
\begin{align}
    \widehat{\Psi}^{[1]}_{3|1}(t) =\int^{t}_0 \left\{ \prod_{(0,u)} \left[1 - \frac{d\widehat{N}_{3|1}(v) + d\widehat{N}_{4|1}(v)}{\widehat{Y}_{0^*|1}(v)}  \right] \right\} \frac{d\widehat{N}_{3|1}(u)}{\widehat{Y}_{0^*|1}(u)}.
\end{align}

\subsection{Extension to general multistate model with a tree structure}\label{sec: general}
Next, we expand the explanation of the estimator for $\Psi_{k|j} = P(X_{k} = 1 ~|~X_{j} = 1)$ to encompass a broader multistate model context. %{\color{black}{It is important to note that the conditional probability, $\Psi_{k|j}$, differs from the probability of transitioning directly from state $j$ to state $k$ within a specific time interval $(s,t)$, represented as $P_{jk}(s,t) = P(S(t) = k\ |\ S(s) = j); ~ s<t$.}} 
The estimator (based on fractional weights) of the probability of occupying state $k$ given a prior visit to state $j$ is denoted as $\widehat{\Psi}^{[1]}_{k|j}$. The estimation of ${\Psi}_{k|j}$ relies on the assumption that state $j$ lies on the unique path connecting states $0$ and $k$; otherwise, ${\Psi}_{k|j} = 0$. Let $m$ represent the number of transitions needed to move from state $0$ to state $k$. Constructing the estimator for $\Psi_{k|j}$ follows a recursive approach based on $m$.

When $m=1$, the computation of $\widehat{\Psi}^{[1]}_{k|j}$ involves using the Aalen-Johansen estimator for transition probabilities within a competing risks framework. On the other hand, when $m=2$, then, one can obtain $\widehat{\Psi}^{[1]}_{k|j}$  by following the procedure outlined in Section \ref{sec: 5state}. More generally, for any state $k$ that is situated at a distance of $m$ from state 0, the estimation of ${\Psi}_{k|j}$ relies on the application of the chain rule of conditional probability, given that the states of interest are aligned along a distinct path originating from the root node. 

%{\color{black}{}}
Following the application of the chain rule, for every state $j$ on the path from state 0 to $k$, where $k$ is a distance $m$ from state 0; define
\begin{align}
    \widehat{\Psi}^{[1]}_{k|j} = \widehat{\Psi}^{[1]}_{\Tilde{k}|j} \cdot \widehat{\Psi}^{[1]}_{k|\Tilde{k}}
    \label{eqn: est1a}
\end{align}
where $\Tilde{k}$ is the state just before state $k$ on the path from $0$ to $k$. From this recursive formula, $\widehat{\Psi}^{[1]}_{\Tilde{k}|j}$ is available since the distance from state 0 to state $\Tilde{k}$ is less than $m$. To construct the estimator $\widehat{\Psi}^{[1]}_{k|\Tilde{k}}$, we utilize the idea of estimating transition probabilities in the competing risk framework and the concept of ``pooling of states''. Thus, we have
\begin{align}
    \widehat{\Psi}^{[1]}_{k|\Tilde{k}} = \int^{\infty}_{0} \widehat{S}_{0^*|\Tilde{k}}(s-) \frac{d\widehat{N}_{k|\Tilde{k}}(s)}{\widehat{Y}_{0^*|\Tilde{k}}(s)}.
    \label{eqn: est1b}
\end{align}
where
\begin{align*}
    \widehat{S}_{0^*|\Tilde{k}}(t) &= \prod_{(0,t)} \left[1 - \frac{d\widehat{N}_{.|\Tilde{k}}(v)}{\widehat{Y}_{0^*|\Tilde{k}}(v)}  \right].
\end{align*}
Here, state $0^*$ denotes an artificial state constructed by combining the states preceding (and including) $\Tilde{k}$. Also,
\begin{align*}
    \widehat{N}_{k|\Tilde{k}}(t) &=  \sum^{n}_{i=1} \left\{\frac{\widehat{N}^P_{\Tilde{k}k}(t) K_h(C_{i} - t)}{\sum^{n}_{i=1} K_h(C_{i} - t)} \right\} \text{ and } \widehat{N}_{.|\Tilde{k}}(t)  = \sum_k \widehat{N}_{k|\Tilde{k}}(t),\\
     \widehat{Y}_{0^*|\Tilde{k}}(t) &= \sum^{n}_{i=1} \left\{\widehat{\phi}_{i\Tilde{k}} \times \frac{\mathcal{I}(S_{i}(C_{i}) \in 0^*) K_h(C_{i} - t)}{n^{-1}\sum^{n}_{i=1} K_h(C_{i} - t)} \right\},
\end{align*}
where $\widehat{N}^P_{\Tilde{k}k}(\cdot)$ is obtained by the isotonic regression of the pairs $(C_i,\ \mathcal{I}(U_{i, \Tilde{k}k} \leq C_i))$. The estimator of the number of persons at risk in state $0^*$, $\widehat{Y}_{0^*|\Tilde{k}}(t)$, is based on the fractional weight $\widehat{\phi}_{i\Tilde{k}}(C_i, S_i(C_i))$ that estimates the probability of individual $i$ reaching state $\Tilde{k}$ based on the information observed at their inspection time $C_{i}$. We interpret $\widehat{\phi}_{i\Tilde{k}}(C_i, S_i(C_i))$ as the contribution of the $i$th individual to the ``at-risk'' set of transitioning out of state $\Tilde{k}$. If individual $i$ is observed to occupy state $\Tilde{k}$ at their inspection time, then, $\widehat{\phi}_{i\Tilde{k}}(C_i, S_i(C_i)) = 1$, otherwise if they were observed to occupy a state not on the path from state 0 to state $\Tilde{k}$, then $\widehat{\phi}_{i\Tilde{k}}(C_i, S_i(C_i)) = 0$. The estimation of $\widehat{\phi}_{i\Tilde{k}}(C_i, S_i(C_i)) = \widehat{\mathbb{P}}(X_{i\Tilde{k}}=1 \ | \ S_i(C_i)= \ell)$ only becomes intricate when such individual occupies a state $\ell$ which lies on the path from state 0 to state $\Tilde{k}$ at their inspection time. To calculate $\widehat{\phi}_{i\Tilde{k}}(C_i, S_i(C_i))$ for such cases, we utilize a competing risk calculation where the states prior of $\Tilde{k}$ are combined to form an event-free state.

Let $T_{k}$ denote the random time at which state $k$ is reached. Estimation of the subdistribution function, $\Psi_{k|j}(t) = \text{Pr}(T_k \leq t,X_k = 1 \mid X_j = 1)$ also follows (\ref{eqn: est1a}). That is; $\widehat{\Psi}^{[1]}_{k|j}(t) = \widehat{\Psi}^{[1]}_{\Tilde{k}|j}(t) \cdot \widehat{\Psi}^{[1]}_{k|\Tilde{k}}(t)$.

{\color{black}We are also interested in the entry distribution for state $k$ given a previous occupancy of state $j$, expressed as 
$${F}_{k|j}(t) = \text{Pr}(T_k \leq t \mid X_k=1,\ X_j=1),$$
that is, the cumulative distribution function of $T_k$ among individuals who eventually reach $k$ after having reached $j$. By conditioning on $X_j=1$, we have
\begin{align*}
\begin{aligned}
        {F}_{k|j}(t) &= \frac{\text{Pr}(T_k \leq t,\ X_k=1 \mid X_j=1)}{\text{Pr}(\ X_k=1 \mid X_j=1)} = \frac{\Psi_{k|j}(t)}{\Psi_{k|j}}
\end{aligned}
\end{align*}
Then, the estimator is obtained by
\begin{align}
        {F}_{k|j}(t) = \frac{\widehat{\Psi}_{k|j}(t)}{\widehat{\Psi}_{k|j}}.
    \label{eqn: entry}
\end{align}
}
% Note that $\widehat{F}_{k|j}(t)$ is the normalized sum of estimated occupation probabilities (conditioned on a prior visit to state $j$) of state $k$ and all other states that come after state $k$ in the progressive system.

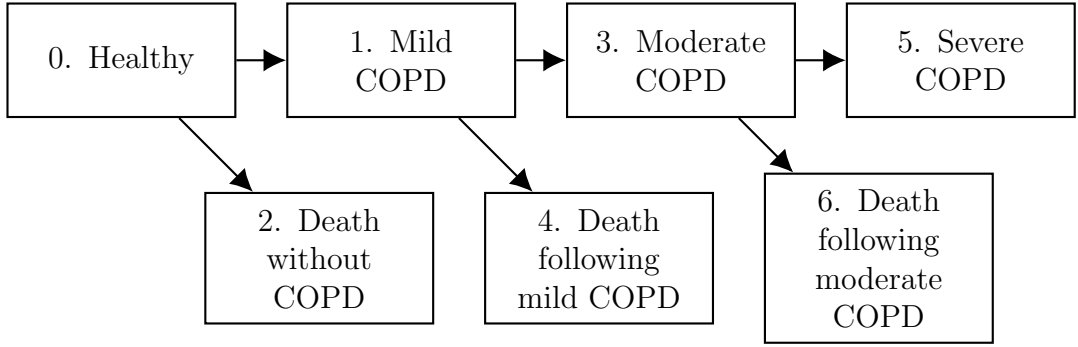
\begin{figure}
\centering
\begin{tikzpicture}[node distance={3.7cm}, thick, state/.style = {draw, rectangle, align=center,inner sep=0.5em, text width=2.6cm,minimum height=1.5cm}] 
\node[state] (0) { 0. Healthy}; 
\node[state] (1) [right of=0] {1. Mild COPD}; 
\node[state] (2) [below right of=0] { 2. Death without COPD}; 
\node[state] (3) [right of=1] { 3. Moderate COPD}; 
\node[state] (4) [below right of=1] { 4. Death following mild COPD}; 
\node[state] (5) [right of=3] { 5. Severe COPD}; 
\node[state] (6) [below right of=3] { 6. Death following moderate COPD}; 
\draw[->] (0) -- (1); 
\draw[->] (0) -- (2); 
\draw[->] (1) -- (3); 
\draw[->] (1) -- (4); 
\draw[->] (3) -- (5); 
\draw[->] (3) -- (6); 
\end{tikzpicture} 
\caption{A seven-state multistate system.}
\label{fig: seven-illness-death}
\end{figure}

For a practical illustration, consider a seven-state multistate system (refer to Figure \ref{fig: seven-illness-death}). Suppose we aim to estimate the probability of occupying state 5 given a prior visit to state 1, denoted by $\Psi_{5|1}$. Utilizing (\ref{eqn: est1a}), in this case, we have $k=5$, $j=1$, and $\widetilde{k}=3$. Consequently, our goal is to compute $\widehat{\Psi}^{[1]}_{5|1} = \widehat{\Psi}^{[1]}_{3|1} \cdot \widehat{\Psi}^{[1]}_{5|3}$. Following the recursive calculation, $\widehat{\Psi}^{[1]}_{3|1}$ is already available. To calculate $\widehat{\Psi}^{[1]}_{5|3}$ using (\ref{eqn: est1b}), define the artificial state $0^* = \{0,1,2,3,4\}$. Moreover, for the non-trivial case, the fractional weight $\widehat{\phi}_{i3}$ is obtained by estimating the probability of transitioning from another artificial state defined by pooling states $\{0,1,2\}$, to state 3.

\section{Product limit estimators}\label{sec: PLE}
%{\color{black}{}}
We introduce a novel estimator for the conditional state occupation probability $\Psi_{k|j}$. This estimator is expressed in terms of marginal state occupation probabilities, which leverage the tree structure of the multistate system. In such a system, any state is reached from the initial state, 0, through a unique path.  Define $\mathcal{S}^{k}$ as the set of states $\ell$ such that state $k$ is on the path from 0 to $\ell$. Then, the conditional quantity, $\Psi_{k|j}$, can be expressed as the ratio of two marginal functions given by
\begin{align}
    \Psi_{k|j}(t) = \frac{\mathbb{P}(S(t) \in \mathcal{S}^k)}{\mathbb{P}(S(\infty) \in \mathcal{S}^j)}
    \label{eqn: est2}
\end{align}
where the numerator is the marginal occupation probability of occupying state $k$ or any of the other states that come after $k$ in the progressive system at time $t$, while the denominator is the marginal occupation probability of occupying state $j$ or any of the other states that come after $j$ in the system during an individual's lifetime. The resulting estimator, denoted by $\widehat{\Psi}^{[2]}_{k|j}$, is constructed from the ratio of these marginal state occupation probabilities.

To construct $\widehat{\Psi}^{[2]}_{k|j}$, we begin by estimating the marginal state occupation probabilities using the product-limit (Aalen-Johansen) formula. This method involves first estimating the counting process and at-risk set, followed by the calculation of the marginal state occupation probabilities. For completeness, we provide the development of these nonparametric estimators under current status censoring in Appendix \ref{sec: datta2006}, following the methodology of Datta and Sundaram \citep{datta2006}. Once these marginal estimates are obtained, the estimator $\widehat{\Psi}^{[2]}_{k|j}$ is derived through a straightforward plug-in procedure.

Following the construction of $\widehat{\Psi}^{[2]}_{k|j}$ using the product-limit approach, the estimation of the conditional entry distribution $F_{k|j}(t)$ becomes straightforward by the substitution of $\widehat{\Psi}^{[2]}_{k|j}$ into (\ref{eqn: entry}).

% To compute $\widehat{\Psi}^{[2]}_{k|j}$, we utilize results from Datta and Sundaram for estimating the marginal state occupation probabilities by expanding the multistate system into a tree structure. First, we estimate the counting process and at-risk set, followed by the estimation of the SOP by the Aalen-Johansen formula. For completeness, we detail the development of these nonparametric estimators under current status censoring, following the approach of Datta and Sundaram \citep{datta2006}, in Appendix \ref{sec: datta2006}. Once these marginal estimates are obtained, $\widehat{\Psi}^{[2]}_{k|j}$ is derived via the plug-in method.

%To compute $\widehat{\Psi}^{[2]}_{k|j}$, we first estimate the marginal state occupation probabilities by utilizing the counting process and at-risk set, followed by obtaining the product-limit (Aalen-Johansen) estimates. For completeness, we detail the development of these nonparametric estimators under current status censoring, following the approach of Datta and Sundaram \citep{datta2006}, in Appendix \ref{sec: datta2006}. Once these marginal estimates are obtained, $\widehat{\Psi}^{[2]}_{k|j}$ is derived via the plug-in method.
 
% To formulate $\widehat{\Psi}^{[2]}_{k|j}$, we use the product-limit approach (see details below) of \cite{datta2006} to estimate the marginal probabilities in the numerator and denominator of (\ref{eqn: est2}). 

\subsection{A toy example}
We demonstrate the formulation of the new estimator using an illustrative multistate system (refer to Figure \ref{fig: seven-illness-death}) that can describe the progression of Chronic Obstructive Pulmonary Disease (COPD). This system comprises seven distinct states representing various health conditions individuals may experience over time. Commencing from the healthy state (alive and event-free), which mirrors normal lung function, individuals might either die without any COPD events or transit to the mild COPD state upon the onset of mild airflow limitation typical in early-stage COPD. Subsequently, from the mild COPD state, individuals could either decease or progress to moderate COPD as their symptoms exacerbate, and airflow limitation intensifies. Advancing from the moderate COPD state, individuals could die or further deteriorate to severe COPD, typified by notable limitations in physical activity, frequent exacerbations, and a heightened risk of respiratory failure. Our focus lies in estimating the proportion of individuals that will reach the severe COPD state given a past occupation of the mild COPD state. This conditional probability is denoted by $\Psi_{5|1}$. Additionally, we will describe the estimation of $\Psi_{3|1}$ and $\Psi_{5|3}$.

The expression in (\ref{eqn: est2}) leads to the respective estimators for these temporal functions:
\begin{align*}
    \widehat{\Psi}^{[2]}_{3|1}(t) &= \frac{\widehat{\pi}_{3^*}(t)}{\widehat{\pi}_{1^*}} = \frac{\widehat{\mathbb{P}}\Big(S(t) \in \{3,5,6\} \Big)}{\widehat{\mathbb{P}}\Big(S(\infty) \in \{1,3,4,5,6\} \Big)} \\
    \widehat{\Psi}^{[2]}_{5|3}(t) &= \frac{\widehat{\pi}_{5^*}(t)}{\widehat{\pi}_{3^*}} = \frac{\widehat{\mathbb{P}}\Big(S(t) \in \{5\} \Big)}{\widehat{\mathbb{P}}\Big(S(\infty) \in \{3,5,6\} \Big)} \\
    \widehat{\Psi}^{[2]}_{5|1}(t) &= \frac{\widehat{\pi}_{5^*}(t)}{\widehat{\pi}_{1^*}} = \frac{\widehat{\mathbb{P}}\Big(S(t) \in \{5\} \Big)}{\widehat{\mathbb{P}}\Big(S(\infty) \in \{1,3,4,5,6\} \Big)}.
\end{align*}
Let's consider the estimators for ${\Psi}_{3|1}(t)$ and ${\Psi}_{5|1}(t)$. For $\widehat{\Psi}_{3|1}(t)$, the numerator estimates the probability of occupying the artificial state $3^*$ that comprises state 3 and other subsequent states along its path, at time $t$. Whereas, for $\widehat{\Psi}_{5|1}(t)$, the numerator estimates the occupation probability of state 5 at time $t$. The denominator of the respective estimators has the same form. Under the competing risk framework, the denominator simply estimates the transition probability of moving from state 0 to state 1 at some time during the observation period. In terms of state occupation probabilities, the denominator is the probability of occupying the artificial state $1^*$ that comprises state 1 and other subsequent states along its path at some time during the observation period.

\begin{figure}[h]
\centering
\includegraphics[scale=0.8]{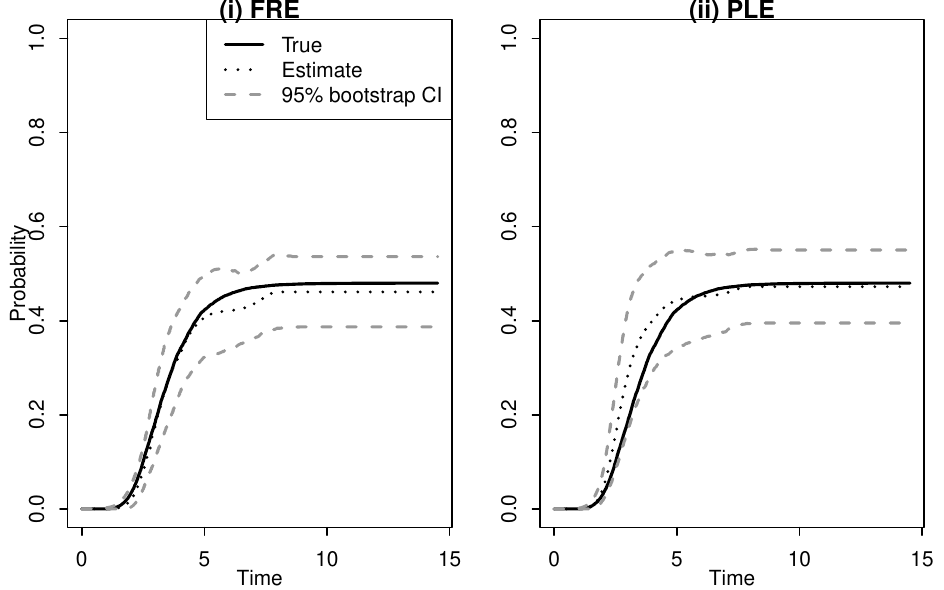}%
\caption{Estimated probability curves for $\Psi_{5|1}(t)$ using the two proposed methods, along with 95\% pointwise bootstrap confidence intervals, based on simulated data from a seven-state model. For $N=1000$ individuals, true transition times were generated from a lognormal distribution, while inspection times followed a uniform distribution.}
\label{fig: P51_fre_ple}
\end{figure}

For a simple comparison of the new approach based on the ratio of product limit estimators (PLE, hereinafter) with the approach based on the fractional risk sets (FRE, hereinafter), Figure \ref{fig: P51_fre_ple} shows the curves from the estimation of the probability of occupying state 5 conditional on a previous visit of state 1, ${\Psi}_{5|1}(t)$, using simulated data. The 95\% pointwise confidence interval (see Section \ref{sec: confint}) corresponding to each probability curve is also shown on the plots. Based on this simulated setting, the plots reveal the validity of both methods in estimating the true probability function. Moreso, the estimators are comparably similar. In Section \ref{sec: sim}, we carry out a thorough investigation comparing the performance of both estimators.

%\FloatBarrier
%\FloatBarrier

\section{Bootstrap-based Confidence Intervals}\label{sec: confint}
The utilization of nonparametric regression in conjunction with the PAV step for formulating the estimators proposed in Section \ref{sec: FRE} and \ref{sec: PLE} imposes challenges for the asymptotic analysis for these estimators. To quantify uncertainty, a bootstrap-based procedure may provide a practical alternative. However, Faraway and Jhun \cite{faraway1990} pointed out that a direct application of the bootstrap will generally lead to invalid results for the ``smoothed'' nonparametric estimators since it cannot capture the bias component of the smoothed estimator.

Following guidelines from existing work on the bootstrap for nonparametric regression estimators \citep{li2001, shao1995}, we describe a smoothed bootstrap procedure for estimating the pointwise confidence intervals for our estimators. {\color{black} Let $\theta(t)$ denote either $\Psi_{k\mid j}(t)$ or $F_{k\mid j}(t)$.} First, using random sampling with replacement, we generate bootstrap censored times $\{C^*_i;\ 1,...,n\}$ from the original data $\{C_i;\ 1,...,n\}$. For the $i$th observation, we then generate the corresponding state $S^*_i(C^*_i)$ from the distribution given by
\begin{align*}
    \text{Prob}(S=j) = \frac{\sum^{n}_{l=1} \mathcal{I}(S_l(C_l)=j) K_{\Tilde{h}}(C_l - C^{*}_{i})}{\sum^{n}_{l=1} K_{\Tilde{h}}(C_l - C^{*}_{i})}.
\end{align*}
In other words, we generate $S^*_i(C^*_i)$ from a smoothed distribution of the states whose inspection times were close to $C^*_i$. Smoothing reduces variance by borrowing strength across nearby estimates, but at the potential cost of increased bias. The bias is mitigated by selecting an appropriate bandwidth using data-driven approaches {\color{black}(see Section \ref{sec: bandwidth})}, as excessively large bandwidths can oversmooth, distorting the probability estimates. Following the theory of the smoothed bootstrap, the bandwidth $\Tilde{h} = h^{\omega}$ needs to be larger than the bandwidth $h$ selected in the estimation based on the original data. We recommend setting $\omega=0.8$ if $0 < h < 1$, and $\omega=1.2$ if $h > 1$.

{\color{black}
Because $\theta(t)$ is a probability-valued function and may lie near 0 or 1, we apply the variance-stabilizing transformation, $g(x)=\sin^{-1}(\sqrt{x})$, which produces a more symmetric and stable sampling distribution. The bootstrap distribution of
\begin{align*}
    \big|g(\widehat{\theta}^*(t; h)\big) -g \big(\widehat{\theta}(t; \Tilde{h}) \big)\big|
\end{align*}
is used to estimate pointwise variability on the transformed scale. For $0<\alpha <1$, let $\widehat{\Delta}_{\alpha}(t)$ denote the $(1-\alpha)$ quantile of this deviation. Here, $\widehat{\theta}^*(t; h)$ is calculated using the bandwidth $h$ from the original estimation but is based on the bootstrap sample, while $\widehat{\theta}(t; \Tilde{h})$ uses the new bandwidth $\Tilde{h}$ but is based on the original sample. The $(1 - \alpha) \times 100 \%$ pointwise CI for $\theta(t)$ is given by 
\begin{align*}
    \Big[g^{-1} \Big\{&\max \Big(0, g\Big(\widehat{\theta}(t; h)\Big) - \widehat{\Delta}_{1-\alpha/2}(t) \Big) \Big\},\ g^{-1} \Big\{\min \Big(\frac{\pi}{2}, g\Big(\widehat{\theta}(t; h)\Big) + \widehat{\Delta}_{1-\alpha/2}(t) \Big) \Big\} \Big],
\end{align*}
where $g^{-1}(x)=\sin^{2}(x)$. The truncation at 0 and $\pi/2$ ensures that the inverse transformation yields valid probability bounds in $[0,1]$. The construction of the confidence interval described above is similar to Datta and Sundaram \cite{datta2006}. In Section \ref{sec: sim}, we conduct a Monte Carlo study to examine the behavior of the smoothed bootstrap confidence intervals.
}

\section{Testing of covariate effects via pseudo-value regression}\label{sec: pseudo}
We use pseudo-value regression to test the effect of baseline covariates, $\mathbf{Z}$, on the distribution of state entry times conditional on a prior state visit. Originally proposed by Andersen et al. \cite{andersen2003} for multistate models with right censoring, this approach has recently been validated by Anyaso-Samuel et al. \cite{anyaso2023} for multistate current status data.

Let $\widehat{\theta}(t)$ denote the marginal estimator of interest, which, in our context, may represent either $\Psi_{k|j}(t)$ or $F_{k|j}(t)$. The jackknife pseudo-values are defined as:
\begin{align}
Y_i(t) = n \widehat{\theta}(t) - (n-1)\widehat{\theta}_{-i}(t),
\end{align}
where $\widehat{\theta}_{-i}(t)$ is calculated by omitting the $i$th subject from the estimation. Each pseudo-value $Y_i(t)$ quantifies the individual contribution of the $i$th subject to the marginal estimator and can be modeled approximately linearly in regression analyses.

To evaluate covariate effects, pseudo-values serve as the response variables in regression models. Specifically, we compute pseudo-values at ten equally spaced points across the observed event-time range. Generalized estimating equations (GEEs) with an independence working correlation structure are then used for inference regarding covariate effects.

We provide detailed steps for implementing pseudo-value regression in Appendix \ref{sec: pseudo2}. Extensive simulation studies, presented in Section A of the Web Supplementary material, validate the performance of this procedure.

\section{Simulation studies}\label{sec: sim}
We performed extensive simulation studies to evaluate the performance of the proposed methods based on: (i) fractional risk sets (FRE), and (ii) the ratio of product limit estimators (PLE) for estimating the conditional quantities of interest. {\color{black}This section focuses on settings without covariate adjustment; results for scenarios with covariates are presented in Section~A of the Web Supplementary Material.} We considered two separate multistate systems: the five-state illness-death model shown in Figure \ref{fig: illness-death}, and the seven-state multistate model shown in Figure \ref{fig: seven-illness-death}. For each multistate system, the exact transition times were generated from a common lognormal hazard. We considered sample sizes of 100, 200, 500, and 1000.

\subsection{Five-State Illness-Death Model}
Earlier, in Section \ref{sec: 5state}, we introduced the five-state illness-death model. For our simulation experiment, our primary interest lies in estimating the probability of occupying state 3 having visited state 1, denoted by $\Psi_{3|1}$. We also study the performance of the estimators for the subdistribution function $\Psi_{3|1}(t)$, and the entry distribution for state 3 given the prior occupation of state 1, denoted by $F_{3|1}(t)$. Our data simulation process initiates with each individual starting in state 0 at time 0, with potential transitions to either state 1 or state 2. Specifically, we simulate the scenario where 60\% of individuals follow the $0 \rightarrow 1$ transition path, while the remaining 40\% follow the $0 \rightarrow 2$ transition path. The transition out of state 0 is governed by an independent Bernoulli distribution. For those individuals transitioning to state 1, we further simulate their progression, with 60\% transitioning to state 3 and 40\% to state 4. This transition from state 1 is also controlled by a separate independent Bernoulli distribution. Following this simulation scheme, the true value of $\Psi_{3|1}$ is 0.6.

Let $V_{i,0}$ and $V_{i,1}$ denote the respective waiting time in states 0 and 1, for individual $i$. We simulate: $V_{i,0} \sim \text{LogNorm}(0, 0.5)$ and $V_{i,1} \sim \text{LogNorm}(0, 0.5)$. We study the performance of our proposed estimators under the scenarios where the random inspection time for each individual is generated from (i) the uniform distribution with a lower bound of 0 and an upper bound equal to the maximum waiting time observed across all individuals in the sample, and (ii) the Weibull distribution with (shape) $\eta = 3$  and (scale) $\tau = 2.5$.

\subsection{Seven-State Multistate Model}
Here, we consider the more complex multistate system shown in Figure \ref{fig: seven-illness-death}. Primarily, we are interested in estimating the probability of occupying State 5 having visited State 1, denoted by $\Psi_{5|1}$. From {\color{black}(\ref{eqn: est1a})}, we have $\Psi_{5|1} = \Psi_{3|1} \times \Psi_{5|3}$. In addition to studying the performance of the estimators for $\Psi_{5|1}$, we will also study the performance of the estimators for $\Psi_{5|3}$, along with the subdistribution functions; $\Psi_{5|1}(t)$ and $\Psi_{5|3}(t)$, and the condition state entry distributions; $F_{5|1}(t)$, and $F_{5|3}(t)$. For the data simulation, all individuals start in state 0 at time 0, and they have two possible paths for transition: either to state 1 or state 2. We generate the data as follows: 80\% of the individuals follow the $0 \rightarrow 1$ path, while the remaining 20\% follow the $0 \rightarrow 2$ path. Among those who transitioned to state 1, there are two possible paths: 70\% transition to state 3, and the remaining 30\% transition to state 4. Moreover, among those that visited state 3, 60\% proceeded to state 5, while the remaining 40\% transitioned to state 6. Note that, the respective transitions out of state 0, state 1, and state 3 are controlled by three separate Bernoulli distributions. Following this simulation approach, the true value of $\Psi_{5|1}$ is 0.6.

To explore the performance of our proposed estimators, we simulate individual waiting times in states 0 and 1 denoted by $V_{i,0}$ and $V_{i,1}$ respectively. We simulate: $V_{i,0} \sim \text{LogNorm}(0, 0.5)$, $V_{i,1} \sim \text{LogNorm}(0, 0.5)$, and $V_{i,3} \sim \text{LogNorm}(0, 0.7)$. We consider two scenarios where we generate the random inspection time for each individual from (i) the uniform distribution with limits 0 and the maximum waiting time across the entire sample and, (ii) the Weibull distribution with $\eta = 2.5$  and $\tau = 4.5$.

\subsection{Simulation results}
We evaluated the performance of the estimator of the conditional state occupation probability $\Psi_{k|j}$ using absolute bias. For the global assessment of the estimators of the subdistribution functions, $\Psi_{k|j}(t)$, and the conditional state entry time distribution, $F_{k|j}(t)$, we employed the mean absolute distance (MAD), defined as follows:

\begin{align}
\text{MAD} = \mathbb{E} \int |\widehat{\theta}^{\text{Emp}}(x) - \widehat{\theta}(x)|\ d\theta_{n}(x),
\label{eqn: MAD}
\end{align}
where $\widehat{\theta}$ represents the estimator based on the current status data, and $\widehat{\theta}^{\text{Emp}}$ corresponds to the empirical estimator based on complete data - this served as the benchmark. To compute the MAD, we used the empirical distribution function of the inspection times, denoted by $\theta_n(x) = \frac{1}{n} \sum_{i=1}^{n} \mathcal{I}(C_i \leq x)$. When the MAD value is equal to 0, it indicates that the estimators are in complete agreement on the support of the observed inspection times. We calculated the absolute bias and MAD via Monte Carlo averaging with a replication size of 1000.

For the five-state illness death model, Table \ref{tab: 5state} shows the results for the estimators constructed with the two proposed methods under the simulation setting where the true transition times follow a lognormal hazard. The results are shown for the separate scenarios where the inspection time for each individual is generated from the uniform and Weibull distribution. For each estimator, the bias and MAD values decrease with increasing sample size, thus suggesting the consistency of the estimators. Although the performance of both methods in estimating the $\Psi_{3|1}$ is comparable in most cases, the estimator based on the FRE approach tends to have a smaller bias. This finding is consistent when we consider the other temporal functions of interest. Generally, the performance of the estimators is reasonable considering the severe censoring of the transition times.

Table \ref{tab: 7state} shows the simulation results for the seven-state multistate system where the true transition times are generated from the lognormal hazard. For each quantity of interest estimated by either the FRE or PLE approach, the performance of the estimators appears to improve with increasing sample size. Under this system, we observe that the estimators based on the FRE method slightly outperform the estimators based on the PLE method. As pointed out in \cite{lan2010a}, a possible explanation of the superiority of the FRE method over the PLE method especially for later states in the system is that the PLE-based estimation of the functions for the later states depends on the estimation done in prior states, therefore, the estimation error may propagate along the path.

{\color{black}
\subsubsection{Coverage of smoothed bootstrap confidence intervals}\label{sec: CI_simres}
For each multistate system considered in this simulation study, we evaluate the performance of the smoothed-bootstrap pointwise confidence intervals. Confidence intervals were obtained using 1000 bootstrap resamples, and empirical coverage probabilities were computed from 1000 Monte Carlo iterations at a nominal level of 0.05. Table~\ref{tab: BS_psi31} reports coverage for the estimators of $\Psi_{3\mid 1}(t)$ under the five-state illness--death model, while Table~\ref{tab: BS_psi51} reports coverage for $\Psi_{5\mid 1}(t)$ under the seven-state model. The results correspond to the setting where inspection times are generated from a uniform distribution.

Because current-status inference at time $t$ is driven primarily by subjects inspected near $t$, coverage can be unstable near boundary regions where the inspection-time density (and hence the effective sample size) is small. We therefore report coverage at five fixed time points chosen away from the boundaries of the inspection-time support. Overall, FRE yields coverage close to the nominal level with mild deviations, whereas PLE tends to be conservative, with coverage often exceeding 0.95. The conservativeness is more pronounced for $\Psi_{5\mid 1}(t)$, reflecting the reduced information available for a state deep in the progressive tree, particularly at later times.
}

\begin{comment}
    
\section{SEER data example}\label{sec: eortc}
\begin{itemize}
    \item Did every patient begin from the early stage cancer?
    \item Check ${\Psi}_{3|1} + {\Psi}_{4|1} = 1$?
\end{itemize}

\begin{figure}
\centering
\begin{tikzpicture}[node distance={3.7cm}, thick, state/.style = {draw, rectangle, align=center,inner sep=0.5em, text width=2.6cm,minimum height=1.5cm}] 
\node[state] (0) { 0. Early}; 
\node[state] (1) [right of=0] {1. Intermediate}; 
\node[state] (2) [below right of=0] {2. Death after early breast cancer}; 
\node[state] (3) [right of=1] {3. Advanced}; 
\node[state] (4) [below right of=1] {4. Death after intermediate breast cancer}; 
\node[state] (5) [right of=3] {5. Metastatic}; 
\node[state] (6) [below right of=3] {6. Death after advanced breast cancer}; 
\node[state] (7) [below right of=5] {7. Death after metastatic breast cancer}; 
\draw[->] (0) -- (1); 
\draw[->] (0) -- (2); 
\draw[->] (1) -- (3); 
\draw[->] (1) -- (4); 
\draw[->] (3) -- (5); 
\draw[->] (3) -- (6); 
\draw[->] (5) -- (7); 
\end{tikzpicture} 
\caption{A seven-state multistate system.}
\label{fig: seer-msm}
\end{figure}
\end{comment}

\section{Application to emulated current status data from a real-world study}\label{sec: eortc}
% Estimating state entry probabilities conditional on prior state occupation is an important problem in many biomedical contexts involving disease progression. However, real-world current status data from complex multistate systems are rarely available in public datasets, despite their practical relevance. 
To illustrate our proposed estimators, we used data from a large randomized clinical trial conducted by the European Organization for Research and Treatment of Cancer (EORTC trial 10854), involving 2,793 patients with early-stage breast cancer. The trial compared outcomes between patients receiving surgery alone versus those receiving surgery plus perioperative chemotherapy. Further details are available in Clahsen et al. \cite{clahsen1996} and Van der Hage et al. \cite{van2001}. For this analysis, we constructed a current status version of the dataset to reflect scenarios where each patient is observed only once at a random inspection time. Such designs are relevant in large-scale population studies and clinical follow-up programs where logistical or ethical considerations limit repeated assessments. Similar approaches have been used in recent work \cite{wu2024} to evaluate methodological robustness under severe censoring. Our goal is to examine how well the proposed estimators perform under these realistic but challenging conditions.

\begin{figure}[ht]
\centering
\begin{tikzpicture}[node distance={3.4cm}, thick, state/.style = {draw, rectangle, align=center,inner sep=0.5em, text width=3.5cm,minimum height=1.7cm}] 
\node[state] (0) {0. Surgery}; 
\node[state] (1) [right =1.5cm of 0] {1. Local recurrence only}; 
\node[state] (2) [below right=0.5cm and 1cm of 0] {2. Distant metastasis only}; 
\node[state] (3) [above right=0.5cm and 1cm of 0] {3. Local recurrence and distant metastasis}; 
\node[state] (4) [below of=0] {4. Death without local recurrence or distant metastasis}; 
\node[state] (5) [above right=0.5cm and 1cm of 1] {5. Distant metastasis following local recurrence}; 
\node[state] (6) [right =1.5cm of 1] {6. Death following Local recurrence}; 
\node[state] (7) [right =2cm of 2] {7. Local recurrence following distant metastasis}; 
\node[state] (8) [below right=0.2cm and 1.3cm of 2] {8. Death following distant metastasis}; 
\draw[->] (0) -- (1); 
\draw[->] (0) -- (2); 
\draw[->] (0) -- (3); 
\draw[->] (0) -- (4); 
\draw[->] (1) -- (5); 
\draw[->] (1) -- (6); 
\draw[->] (2) -- (7); 
\draw[->] (2) -- (8); 
\end{tikzpicture} 
\caption{Multistate system for the breast cancer study from EORTC-trial 10854.} 
\label{fig: rott-msm}
\end{figure}
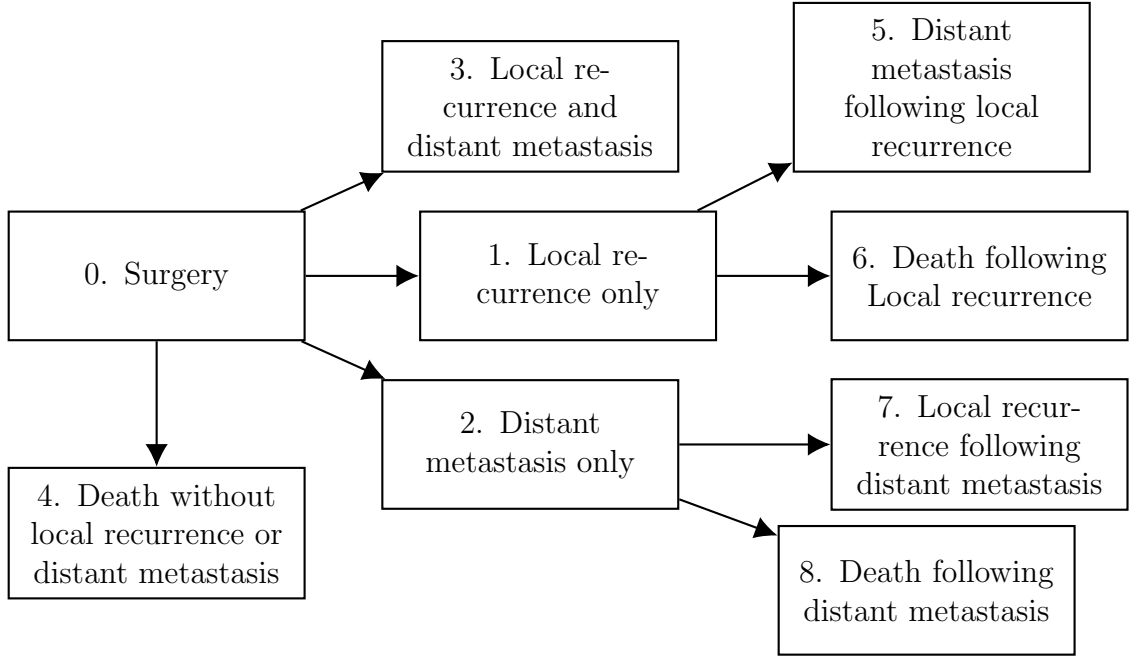

Figure \ref{fig: rott-msm} outlines the progressive nine-state system used in this analysis, with surgery as the time origin (state 0). Subsequent states represent key clinical events, including loco-regional recurrence (state 1), distant metastasis (state 2), simultaneous relapse and metastasis (state 3), and various death states depending on the patient's prior path. Follow-up times ranged from 12 days to 18.5 years, with a median of 10.5 years.

To emulate current status data, each patient was assigned a single inspection time drawn uniformly between 0 and 18.5 years. The state occupied at that inspection time was recorded. If the patient was censored before the assigned inspection time, the most recently observed state was used. This framework represents a severe form of censoring, where only the state at a single, possibly late, time point is known. Table \ref{tab: eortc-states} summarizes the distribution of observed states at inspection. As expected, a large proportion of patients remained in the initial state (alive post-surgery) at inspection.

We focused on estimating the probability of occupying state 5 (distant metastasis), conditional on a prior visit to state 1 (loco-regional recurrence only). This probability, denoted $\Psi_{5|1}$, is of clinical interest for identifying patients at increased risk of progression following recurrence \citep{vicini2003, murata2023}.

The estimates and corresponding 95\% bootstrap confidence intervals based on the FRE and PLE approach are $0.400\ (0.241, 0.570)$ and $0.433\ (0.222,  0.658)$, respectively. The estimation results from both proposed methods are comparable. For context, a marginal (unconditional) analysis estimated the probability of occupying state 5 as only 0.050, emphasizing how conditioning on prior recurrence substantially changes the estimated risk. We also applied the PLE approach to the original right-censored data, yielding an estimate of 0.344 with a 95\% CI of (0.274, 0.422), based on 1,000 bootstrap replications. The proximity of this estimate to those obtained from the current status data underscores the validity and practical utility of our methods, even under severe censoring.

Finally, using the pseudo-value approach, we also tested the mean effect of key prognostic factors, including age, tumor size ($\leq 2$cm, 2-5cm, $>5$cm), nodal status (negative vs. positive), type of surgery (mastectomy with radiation therapy, mastectomy without radiation therapy, or breast conserving), perioperative chemotherapy (yes/no) and adjuvant chemotherapy (yes/no). Initial estimates for the pseudo-value regression were obtained using either the FRE or PLE method. As summarized by Putter et al. \cite{putter2006}, these prognostic factors provide important context for modeling progression. Notably, breast-conserving surgery, compared to mastectomy with radiation therapy, was significantly associated with an increased probability of transitioning to the distant metastasis state following a prior visit to the local recurrence state (FRE $p$-value: 0.022; PLE $p$-value: 0.018).

\section{Discussion}\label{sec: discussion}
This paper carefully employs techniques from the framework of competing risks to construct nonparametric estimators of state entry probabilities and state entry distributions, predicated on the prior occupation of a given state, utilizing current status multistate data. The severe censoring of the event history data posed by a single observation of individuals at different random inspection times brings about additional complexities when attempting to construct nonparametric estimates of temporal functions within a multistate model. Such data are frequently encountered in low-resource settings, where limited follow-up data make it difficult to track disease progression comprehensively. In such scenarios, identifying patients at higher risk of advancing to more severe disease stages is crucial for prioritizing interventions and optimizing resource allocation.

We tackle this problem by harnessing ideas from nonparametric regression modeling to formulate estimators of the counting process and at-risk sets that serve as foundational elements in constructing estimators of the probability functions. To estimate the conditional quantities in the context of current status data, we employed the fraction-risk estimation method that was only available for right-censored data, and we also introduced a novel method based on product-limit estimation. Using data from a real-world application and simulation studies for illustrative purposes, we show that both estimation methods yield valid and comparable results.

In practice, each estimation method has its benefits and challenges. {\color{black}Using the PLE, estimation of $\Psi_{k\mid j}$ depends on estimated marginal state-occupation probabilities for successive states. Consequently, any estimation error in the ``starting’’ marginal SOPs can propagate forward and affect the final estimate. For the FRE approach, if the true fractional weights were known, it would be optimal; in practice, however, these weights must also be estimated recursively, so estimation error may propagate there as well. Computationally, the PLE is easier to implement once the artificial states are specified and marginal SOPs are available from standard software \cite{anyaso2023}. The FRE approach is more involved because it requires calculating fractional weights and combining information across a sequence of previous states. Nonetheless, in our simulations, FRE generally performed better, showing smaller bias in smaller samples and for transitions that occur deeper in the multistate tree.}

{\color{black}
The smoothed-bootstrap pointwise confidence intervals exhibit stable and reliable coverage across a wide range of settings, with slightly wider intervals observed for targets involving states deep in the progressive tree and at later times, where current-status sampling provides less information and the effective sample size is smaller. The PLE intervals are typically wider than those from FRE, reflecting additional variability induced by the ratio-of-marginals construction. For a few early time points, mild undercoverage can occur due to finite-sample skewness near the edge of the time range. Overall, the results highlight that current-status inference for late-state probabilities requires adequate inspection-time support and typically larger sample sizes; further improvements may be obtained by time-adaptive smoothing or more targeted bootstrap calibration.
}

The intuitive approaches we propose provide a more direct route for analyzing the severely censored event data since they utilize the averaging principle rather than maximizing a likelihood function. Suppose we consider a less complicated setting (survival model with two states); the nonparametric maximum likelihood estimator (NPMLE) of the survival function has undesirable properties since it may be undefined over a set of regions. For the general multistate model with current status data, an NPMLE approach would be more problematic.

We operate under the assumption of a progressive multistate system, structured as a directed tree, where each state is reachable via a unique path originating from the root node. This implies knowledge of path information from past states at a random inspection time, although the exact transition times remain unknown. While this assumption often holds due to the inherent structure of the multistate system, there are instances where distinctions among paths can be made by incorporating auxiliary states into an expanded multistate system. Additionally, a considerable number of inspections in a specific state are necessary to calculate the occupation probability of that state. In general, a relatively large sample size is required for valid nonparametric estimation of occupation probabilities, especially when dealing with current status data, where information is significantly limited compared to complete or right-censored data.

We could extend the proposed methods to multistate models with case-II interval-censored data by utilizing similar formulas for estimating the counting process and the at-risk sets. Since the estimators of these functionals are based on averaging, we could sum over the multiple inspection times for each individual. Ignoring the correlation among the inspection times will not inhibit the unbiasedness and consistency properties of the resulting estimators. However, the efficiency of the estimators could be improved by including a weighted variance-covariance matrix in the PAV step of the isotonic regression to account for the events recorded multiple times for each individual.

\section*{Data Availability Statement}
The data from the breast cancer study used in Section \ref{sec: eortc} are available from the European Organization for Research and Treatment of Cancer (EORTC). Restrictions apply to the availability of these data, which were used under license in this paper. 

% \newpage
%\bibliographystyle{chicago}
% \bibliography{Bibliography}

\clearpage
\begin{table}[htb]
\caption{Absolute bias ($\Psi_{3|1}$) and MAD ($\Psi_{3|1}(t)$ \& $F_{3|1}(t)$) for the scenario where the true transition times for the five-state illness-death model were generated from the lognormal distribution. Inspection time for each individual was generated from (i) uniform and (ii) Weibull distribution. The results are shown by the methods proposed: (a) FRE, and (b) PLE. The results are based on a Monte Carlo sample size of 1000.}
\label{tab: 5state}
\centering
\begin{tabular}[t]{lrllllll}
\toprule
\multicolumn{2}{l}{ } & \multicolumn{2}{c}{$\Psi_{3|1}$} & \multicolumn{2}{c}{$\Psi_{3|1}(t)$} & \multicolumn{2}{c}{$F_{3|1}(t)$} \\
\cmidrule(l{3pt}r{3pt}){3-4} \cmidrule(l{3pt}r{3pt}){5-6} \cmidrule(l{3pt}r{3pt}){7-8}
Inspection time & $N$ & FRE & PLE & FRE & PLE & FRE & PLE\\
\midrule
 & 100 & 0.057 & 0.065 & 0.055 & 0.059 & 0.065 & 0.061\\

 & 200 & 0.039 & 0.041 & 0.040 & 0.042 & 0.049 & 0.046\\

 & 500 & 0.026 & 0.025 & 0.029 & 0.029 & 0.035 & 0.032\\

\multirow[t]{-4}{*}{\raggedright\arraybackslash Uniform} & 1000 & 0.019 & 0.018 & 0.023 & 0.022 & 0.027 & 0.025\\
\cmidrule{1-8}
 & 100 & 0.047 & 0.060 & 0.053 & 0.065 & 0.082 & 0.087\\

 & 200 & 0.036 & 0.045 & 0.041 & 0.047 & 0.070 & 0.072\\

 & 500 & 0.028 & 0.030 & 0.029 & 0.032 & 0.054 & 0.054\\

\multirow[t]{-4}{*}{\raggedright\arraybackslash Weibull} & 1000 & 0.025 & 0.025 & 0.023 & 0.024 & 0.046 & 0.045\\
\bottomrule
\end{tabular}
\end{table}

\clearpage
\begin{table}[htb]
\caption{Absolute bias ($\Psi_{5|1}\ \& \ \Psi_{5|3}$) and MAD ($\Psi_{5|1}(t),\ \Psi_{5|3}(t),\ F_{5|1}(t)\ \& \ F_{5|3}(t)$) for the scenario where the true transition times for the seven-state model were generated from the lognormal distribution. Inspection time for each individual was generated from (i) uniform and (ii) Weibull distribution. The results are shown by the methods proposed: (a) FRE, and (b) PLE.  The results are based on a Monte Carlo sample size of 1000.}
\label{tab: 7state}
\centering
\resizebox{1.0\textwidth}{!}{%
\begin{tabular}[t]{lrllllllllllll}
\toprule
\multicolumn{2}{c}{ } & \multicolumn{2}{c}{$\Psi_{5|1}$} & \multicolumn{2}{c}{$\Psi_{5|3}$} & \multicolumn{2}{c}{$\Psi_{5|1}(t)$} & \multicolumn{2}{c}{$\Psi_{5|3}(t)$} & \multicolumn{2}{c}{$F_{5|1}(t)$} & \multicolumn{2}{c}{$F_{5|3}(t)$} \\
\cmidrule(l{3pt}r{3pt}){3-4} \cmidrule(l{3pt}r{3pt}){5-6} \cmidrule(l{3pt}r{3pt}){7-8} \cmidrule(l{3pt}r{3pt}){9-10} \cmidrule(l{3pt}r{3pt}){11-12} \cmidrule(l{3pt}r{3pt}){13-14}
Inspection time & $N$ & FRE & PLE & FRE & PLE & FRE & PLE & FRE & PLE & FRE & PLE & FRE & PLE\\
\midrule
 & 100 & 0.052 & 0.060 & 0.059 & 0.068 & 0.044 & 0.049 & 0.049 & 0.056 & 0.060 & 0.059 & 0.057 & 0.059\\

 & 200 & 0.037 & 0.043 & 0.042 & 0.050 & 0.035 & 0.039 & 0.039 & 0.045 & 0.049 & 0.048 & 0.045 & 0.048\\

 & 500 & 0.024 & 0.027 & 0.027 & 0.031 & 0.026 & 0.029 & 0.028 & 0.034 & 0.038 & 0.040 & 0.033 & 0.040\\

\multirow[t]{-4}{*}{\raggedright\arraybackslash Uniform} & 1000 & 0.018 & 0.019 & 0.021 & 0.022 & 0.021 & 0.025 & 0.022 & 0.029 & 0.030 & 0.035 & 0.025 & 0.035\\
\cmidrule{1-14}
 & 100 & 0.054 & 0.056 & 0.068 & 0.073 & 0.045 & 0.046 & 0.054 & 0.058 & 0.074 & 0.072 & 0.072 & 0.072\\

 & 200 & 0.036 & 0.044 & 0.040 & 0.048 & 0.034 & 0.034 & 0.038 & 0.040 & 0.060 & 0.060 & 0.058 & 0.060\\

 & 500 & 0.027 & 0.031 & 0.030 & 0.034 & 0.026 & 0.023 & 0.028 & 0.028 & 0.049 & 0.048 & 0.047 & 0.048\\

\multirow[t]{-4}{*}{\raggedright\arraybackslash Weibull} & 1000 & 0.022 & 0.023 & 0.025 & 0.027 & 0.023 & 0.018 & 0.022 & 0.022 & 0.044 & 0.041 & 0.038 & 0.041\\
\bottomrule
\end{tabular}
}
\end{table}

\clearpage
\begin{table}[htb]
\caption{Coverage probabilities for the estimator of $\Psi_{3\mid 1}(t)$ under the five-state illness--death model, for the two proposed methods: (i) FRE and (ii) PLE. The confidence intervals are based on 1000 bootstrap samples at a nominal level of 0.05, and the coverage results are based on 1000 Monte Carlo iterations.}
\label{tab: BS_psi31}
\centering
\begin{tabular}[t]{rllllllll}
\toprule
\multicolumn{1}{c}{ } & \multicolumn{4}{c}{FRE} & \multicolumn{4}{c}{PLE} \\
\cmidrule(l{3pt}r{3pt}){2-5} \cmidrule(l{3pt}r{3pt}){6-9}
\multicolumn{1}{c}{ } & \multicolumn{4}{l}{Sample size} & \multicolumn{4}{l}{Sample size} \\
Time & 100 & 200 & 500 & 1000 & 100 & 200 & 500 & 1000\\
\midrule
1.21 & 0.926 & 0.955 & 0.967 & 0.976 & 0.887 & 0.928 & 0.956 & 0.982\\
1.34 & 0.923 & 0.942 & 0.948 & 0.957 & 0.902 & 0.933 & 0.966 & 0.985\\
1.48 & 0.932 & 0.934 & 0.942 & 0.946 & 0.927 & 0.950 & 0.975 & 0.987\\
1.61 & 0.940 & 0.937 & 0.941 & 0.939 & 0.953 & 0.954 & 0.987 & 0.989\\
1.74 & 0.954 & 0.943 & 0.941 & 0.942 & 0.968 & 0.971 & 0.996 & 0.990\\
\bottomrule
\end{tabular}
\end{table}

\clearpage

\begin{table}[htb]
\caption{Coverage probabilities for the estimator of $\Psi_{5|1}(t)$ under the seven-state model, for the two proposed methods: (i) FRE and (ii) PLE. The confidence intervals are based on 1000 bootstrap samples at a nominal level of 0.05, and the coverage results are based on 1000 Monte Carlo iterations.}
\label{tab: BS_psi51}
\centering
\begin{tabular}[t]{rllllllll}
\toprule
\multicolumn{1}{c}{ } & \multicolumn{4}{c}{FRE} & \multicolumn{4}{c}{PLE} \\
\cmidrule(l{3pt}r{3pt}){2-5} \cmidrule(l{3pt}r{3pt}){6-9}
\multicolumn{1}{c}{ } & \multicolumn{4}{l}{Sample size} & \multicolumn{4}{l}{Sample size} \\
Time & 100 & 200 & 500 & 1000 & 100 & 200 & 500 & 1000\\
\midrule
3.89 & 0.969 & 0.950 & 0.948 & 0.935 & 0.993 & 0.989 & 0.996 & 0.982\\
4.05 & 0.979 & 0.957 & 0.961 & 0.951 & 0.992 & 0.990 & 0.995 & 0.987\\
4.21 & 0.987 & 0.965 & 0.970 & 0.967 & 0.991 & 0.990 & 0.995 & 0.993\\
4.37 & 0.991 & 0.974 & 0.977 & 0.973 & 0.992 & 0.992 & 0.997 & 0.994\\
4.53 & 0.993 & 0.982 & 0.983 & 0.977 & 0.994 & 0.995 & 0.997 & 0.994\\
\bottomrule
\end{tabular}
\end{table}

\clearpage
\begin{table}[htb]
\caption{Distribution of state occupation at inspection based on the current status data from the EORTC-trial 10854.}\centering
\label{tab: eortc-states}
\begin{tabular}[t]{llllllllll}
\toprule
State & 0 & 1 & 2 & 3 & 4 & 5 & 6 & 7 & 8\\
\midrule
Frequency & 1941 & 119 & 122 & 66 & 93 & 57 & 29 & 30 & 336\\
Percent & 0.695 & 0.043 & 0.044 & 0.024 & 0.033 & 0.020 & 0.010 & 0.011 & 0.120\\
\bottomrule
\end{tabular}
\end{table}

\appendix
\section{Nonparametric estimators for a multistate system}\label{sec: datta2006}

Recall that $U_{j j'}$ denotes the (unobserved) transition time at which an individual transitions from $j$ to $j'$. To estimate temporal functions of the multistate system, two essential functionals are the counting process denoted by $N_{j j'}(t)$ and the at-risk process denoted by $Y_{j}(t)$. The counting process $N_{j j'}(t)$ counts the number of transitions from state $j$ to $j'$ in the time interval $[0, t)$. On the other hand, the at-risk process $Y_{j}(t)$ represents the number of individuals who are at risk of transitioning out of state $j$ at time $t$. These processes are expressed as follows
\begin{align*}
    N_{j j'}(t) &= \sum^{n}_{i=1} \mathcal{I}(U_{i,j j'} \leq t), \quad \text{ and }\\
    Y_j(t) &= \sum^{n}_{i=1} \mathcal{I}(S_{i}(t-) = j),
\end{align*} 
where $\mathcal{I}(\cdot)$ denotes the indicator function and $S(t-)$ is the state occupied just before time $t$. By the law of large numbers,
\begin{align*}
    n^{-1}N_{j j'}(t) \xrightarrow{P} \text{Pr}(U_{j j'} \leq t),
\end{align*} 
for any $t \geq 0$. Since the inspection time $C$ is independent of the multistate process, we can write 
\begin{align*}
    \text{Pr}(U_{j j'} \leq t) = \mathbb{E}(\mathcal{I}(U_{j j'} \leq C)\ |\ C = t),
\end{align*} 
where $\mathcal{I}(U_{j j'} \leq C)$ is the indicator of the event that the transition from state $j$ to $j'$ has taken place by time $C$. 

Using the theory of nonparametric regression, Datta and Sundaram \cite{datta2006} proposed nonparametric estimators of $N_{j j'}(t)$ and $Y_{j}(t)$ under this context of current status data. For $N_{jj'}(t)$, they proposed a two-step estimation approach where first, isotonic regression via the pooled adjacent violators (PAV) algorithm \citep{barlow1972} is performed based on the pairs $(C_{i},\ \mathcal{I}(U_{i,jj'} \leq C_{i}))$. Let $\widehat{N}^P_{j j'}(t)$ denote the fitted responses from the isotonic regression. Then secondly, kernel smoothing \citep{mukerjee1988, nadaraya1964, watson1964} with a log-concave density, $K>0$ is performed to remove long flat parts of $\widehat{N}^P_{jj'}(t)$. The resulting estimator of $N_{j j'}(t)$ is given by
\begin{align} 
    \widehat{N}_{j j'}(t) =  \sum^{n}_{i=1} \left\{\widehat{N}^P_{j j'}(t) \times \frac{ K_h(C_{i} - t)}{\sum^{n}_{i=1} K_h(C_{i} - t)} \right\},
    \label{eqn: CPest}
\end{align} 
where $K_h(\cdot) = h^{-1}K(\cdot/h)$ with a bandwidth sequence $0 < h = h(n) \downarrow 0$ chosen by the criteria specified by Wand and Jones\cite{wand1994} {\color{black}(see Section \ref{sec: bandwidth})}.
Moreover, to estimate $\text{Pr}(X(t -) = j)$, which represents the probability limit of $n^{-1}Y_{j}(t)$, Datta and Sundaram \cite{datta2006} employed the kernel smoothing approach. Accordingly, the estimation of $Y_{j}(t)$ is given by:
\begin{align} 
    \widehat{Y}_{j}(t) = \sum^{n}_{i=1} \left\{\mathcal{I}(S_{i}(C_{i}) = j) \times \frac{K_h(C_{i} - t)}{n^{-1}\sum^{n}_{i=1} K_h(C_{i} - t)} \right\},
    \label{eqn: ARest}
\end{align} 
where it is assumed that $S_{i}(C_{i} -) = S_{i}(C_{i})$ with probability 1. The estimators in (\ref{eqn: CPest}) and (\ref{eqn: ARest}) allow for inference in multistate models with current status data. Datta and Sundaram \cite{datta2006} further demonstrated their utility in estimating key marginal quantities like transition and state occupation probabilities, based on the product-limit method of Andersen et al. \cite{andersen1993}.

The product-limit estimator of the marginal state occupation probability relies on the Nelson-Aalen estimate of the cumulative transition intensity, denoted as $\boldsymbol{A}(t)$. Let $\widehat{\boldsymbol{A}}(t) = \{\widehat{A}_{jj'}(t)\}$ denote the corresponding estimator given by the expression:
\begin{align} 
        \widehat{A}_{jj'}(t) = 
    \begin{dcases}
    \int^t_0 \mathcal{I} \Big(\widehat{Y}_{j}(u)>0 \Big) \widehat{Y}_{j}(u)^{-1} \text{d}\widehat{N}_{jj'}(u) & j \neq j',\\
    - \sum_{j' \neq j} \widehat{A}_{jj'}(t) & j = j',
    \end{dcases}
    \label{eqn: NelsonAalen}
\end{align} 
where $\widehat{N}_{jj'}(t)$ and $\widehat{Y}_{j}(t)$ respectively denote the estimators of the counting process and the at-risk set (\textit{ref:} Eqns (\ref{eqn: CPest})-(\ref{eqn: ARest})). The product limit of $\widehat{\boldsymbol{A}}(t)$ leads to the (marginal) Aalen-Johansen estimator of the transition probability matrix ${\boldsymbol{P}}(s,t)$,
\begin{align} 
    \widehat{\boldsymbol{P}}(s,t) = \prod_{(s,t]} \{\boldsymbol{I} + \text{d}\widehat{\boldsymbol{A}}(u) \},
    \label{eqn2: transProb}    
\end{align} 
where $\boldsymbol{I}$ denotes the $Q \times Q$ identity matrix. Consequently, the marginal estimator of the state occupation probability is given by
\begin{align} 
    \widehat{\pi}_j(t) = \sum^Q_{j'=1} \widehat{\pi}_{j'}(0) \widehat{P}_{j' j}(0, t),
   \label{eqn: stocc}
\end{align} 
where $\widehat{\pi}_{j'}(0) = \{\widehat{Y}_{j'} (0+)\}/\{\sum^Q_{j=1} \widehat{Y}_j (0+)\}$, and $\widehat{P}_{j' j}(0, t)$ is the $j' j$th element of $\widehat{\boldsymbol{P}}(0,t)$. Datta and Satten \cite{datta2001} demonstrated that (\ref{eqn: stocc}) provides a consistent estimator for the state occupation probability, even in non-Markov models. This finding allows us to estimate the conditional state occupation probability without the Markov assumption using a product-limit approach similar to the Aalen-Johansen calculation.

{\color{black}
\subsection{Bandwidth selection}\label{sec: bandwidth}
Following earlier work on nonparametric estimation for multistate current-status data \citep{datta2006,datta2009}, we adopt a data-driven approach for selecting the bandwidths used in the kernel-smoothing step of our estimators and in the smoothed bootstrap. For consistency of the kernel estimators, the bandwidth $h$ must satisfy the usual conditions: $h\rightarrow 0$, $nh^2 \rightarrow \infty$, $nh^r \rightarrow 0$, for some $r>2$; which balance bias and variance. Datta et al. \citep{datta2009} recommend using a common bandwidth based on the design distribution of the inspection times. In particular, each smoothed process is obtained using a normal kernel with a bandwidth chosen to be asymptotically $L_2$-optimal for estimating the density of the inspection times $C_i$, following the direct plug-in described in Wand and Jones \citep[Section 3.2]{wand1994}. Practically, this bandwidth can be computed using the \texttt{dpik} function in the KernSmooth R package.

In the simulation studies, this procedure yielded bandwidths in the range 0.34–0.59 on the simulated time scales. In the real-data application, where inspection times ranged from 0.83 to 6755.93 days, the selected bandwidth was $h=246.8$ days ($\approx 0.68$ years).
}

\section{Testing covariate effects}\label{sec: pseudo2}
We aim to test the association between covariates and a temporal function, $\theta(t)$, by using pseudo-values as response variables in regression models.

Let $\boldsymbol{Y}_i = \{Y_i(t_1), \ldots, Y_i(t_r)\}$ denote the vector of pseudo-values calculated at time points $(t_1, \ldots, t_r)$. {\color{black}To investigate the relationship between covariates and the temporal function, we assume the following model:}
\begin{align}
g(\theta(t_r \mid \mathbf{Z}_i)) = \alpha_r + \boldsymbol{\zeta}\mathbf{Z}_i,
\label{eqn: pseudo-mod}
\end{align}
where $g(\cdot)$ is an appropriate link function, $\mathbf{Z}_i$ is the vector of baseline covariates, and $\alpha_r$ allows a unique intercept for each time point $t_r$. The vector $\boldsymbol{\zeta}$ quantifies the effects of covariates on the temporal function. {\color{black}The pseudo-value regression model (\ref{eqn: pseudo-mod}) is used as a working marginal model for $\theta(t)$. It need not coincide with the full data-generating mechanism; inference focuses on testing whether covariates have any effect on $\theta(t)$ in this marginal sense.} Typically, pseudo-values are recommended to be computed at a minimum of five equally spaced time points spanning the range of event times \cite{klein2005, andersen2010, anyaso2023}.

We utilize the generalized estimating equation \cite{liang1986} to estimate $\boldsymbol{\beta} = (\alpha_1, \ldots, \alpha_r, \boldsymbol{\zeta})$. We solve the following estimating equation
\begin{align*}
\sum_{i=1}^{m} \boldsymbol{U}_i(\boldsymbol{\beta}, \alpha) = \boldsymbol{0},
\label{eqn: estimating}
\end{align*}
where
\begin{align*}
\boldsymbol{U}_i(\boldsymbol{\beta}, \alpha) &= \boldsymbol{D}_i^T \boldsymbol{V}_i^{-1}(\boldsymbol{Y}_i - g^{-1}(\boldsymbol{\beta}\mathbf{Z}_i)), \\
\boldsymbol{D}_i &= \frac{\partial g^{-1}(\boldsymbol{\beta}\mathbf{Z}_i)}{\partial \boldsymbol{\beta}},
\end{align*}
and $\boldsymbol{V}_i = \sigma^2 \boldsymbol{R}_i(\alpha)$ is the working covariance matrix with correlation structure $\boldsymbol{R}_i(\alpha)$ and dispersion parameter $\sigma^2$.

We estimate the variance of $\widehat{\boldsymbol{\beta}}$ using the robust sandwich estimator:
\begin{align*}
\widehat{\text{Var}}(\widehat{\boldsymbol{\beta}}) = \boldsymbol{B}^{-1} \boldsymbol{M} \boldsymbol{B}^{-1},
% \label{eqn: sandwich}
\end{align*}
where
\begin{align*}
\boldsymbol{B} &= \sum_{i=1}^{m} \boldsymbol{D}_i^T \widehat{\boldsymbol{V}}_i^{-1}\boldsymbol{D}_i, \\
\boldsymbol{M} &= \sum_{i=1}^{m} \boldsymbol{U}_i(\widehat{\boldsymbol{\beta}}, \widehat{\alpha}) \boldsymbol{U}_i(\widehat{\boldsymbol{\beta}}, \widehat{\alpha})^T,
\end{align*}
and $\widehat{\boldsymbol{V}}_i = \widehat{\sigma}^2 \boldsymbol{R}_i(\widehat{\alpha})$. Here, $\widehat{\sigma}$ and $\widehat{\alpha}$ represent consistent estimates of the dispersion parameter and correlation structure parameter, respectively.

\end{document}